# Computationally-efficient Expressions for the Collision Efficiency Between Electrically Charged Aerosol Particles and Cloud Droplets

S.N. Tripathi[1*], S. Vishnoi[1], S. Kumar[1], and R.G. Harrison[2]

[1]Indian Institute of Technology, Kanpur, India
[2]Department of Meteorology, The University of Reading, UK

## Abstract

A multifactor parameterization is described to permit the efficient calculation of collision efficiency ($E$) between electrically charged aerosol particles and neutral cloud droplets in numerical cloud and climate models. The four parameter representation summarizes the results obtained from a detailed microphysical model of collision efficiency, which accounts for the different forces acting on the aerosol in the path of falling cloud droplets. The parameterization's range of validity is for aerosol particle radius 0.4 to 10 μm, aerosol particle density 1 to 2.0 g.cm$^{-3}$, aerosol particle charge from neutral to 100 elementary charges and drop radii 18.55-142 μm. It yields collision efficiencies well within an order of magnitude of the detailed model's values, from a data set of of 3978 $E$ values. 95% of these values have modeled to parameterized ratios between 0.5 and 1.5 for aerosol particle size range 0.4 to 2 μm and about 96 % in the second size range. This parameterization speeds up the collision efficiency calculation by a factor of ~ $10^3$, as compared with the original microphysical model, permitting the inclusion of electric charge effects in numerical cloud and climate models. In the following pages parameterization code in C language is provided for readymade use.

[1] * Corresponding author: Present address: Department of Civil Engineering, IIT Kanpur.

*E-mail address*: snt@iitk.ac.in (S.N. Tripathi)

```
// C  program to calculate collision efficiency for particle range 2 to 10 μm
// checked in Turbo C++ (MS Windows (3.1))
#include <stdio.h>
#include <conio.h>
#include <math.h>
#include <stdlib.h>
#include <iostream.h>

void main()
{
//  p: Aerosol Density(in gm cm-3)
// a: Aerosol Radius(in micrometer)
// A: Drop Radius(in μm)
// q: Aerosol Charge(in e)
// CE: Collision Efficiency
double p;
double a;
double A;
double q;
double CE;
cout<<"Density = ";
cin>>p;
cout<<"\n";
cout<<"Aerosol particle radius(2 to 10) = ";
cin>>a;
cout<<"\n";
cout<<"drop radius = ";
cin>>A;
cout<<"\n";
cout<<"aerosol charge = ";
cin>>q;
cout<<"\n";

            CE =exp(-5.716489195059299e9 - 1.3502467167070127e9*a -
3.756681818771443e7*pow(a,2) +
           1.0274887973391494e7*pow(a,3) - 474581.26517530665*pow(a,4) +
113008.21977161539*pow(a,5) -
           72.46901157193014*pow(a,6) + 1.0794202979739191e10*pow(A,0.004) +
3.512007294785022e9*a*pow(A,0.004) +
           2.3688142700432178e7*pow(a,2)*pow(A,0.004) -
2.2162749340724096e7*pow(a,3)*pow(A,0.004) +
           682166.5339240788*pow(a,4)*pow(A,0.004) -
332633.2975421664*pow(a,5)*pow(A,0.004) +
           142.39869256481296*pow(a,6)*pow(A,0.004) +
9.753162531386845e7*pow(A,0.008) -
```

7.854765376964403e8*a*pow(A,0.008) +
4.658226453561119e7*pow(a,2)*pow(A,0.008) +
6.143561215065787e6*pow(a,3)*pow(A,0.008) +
706262.7385939864*pow(a,4)*pow(A,0.008) +
326425.879615941*pow(a,5)*pow(A,0.008) -
69.97516122969726*pow(a,6)*pow(A,0.008) -
7.695904424515492e9*pow(A,0.012) -
3.1077122146049566e9*a*pow(A,0.012) +
2.0163661339104705e7*pow(a,2)*pow(A,0.012) +
1.387236022204568e7*pow(a,3)*pow(A,0.012) -
1.5648414522853415e6*pow(a,4)*pow(A,0.012) -
106798.08516595581*pow(a,5)*pow(A,0.012) -
2.170233888602743e9*pow(A,0.016) -
3.2724710680248666e8*a*pow(A,0.016) -
3.749353232820472e7*pow(a,2)*pow(A,0.016) -
8.82285425161321e6*pow(a,3)*pow(A,0.016) +
650932.0782810887*pow(a,4)*pow(A,0.016) +
7.2465267264062395e9*pow(A,0.02) +
3.688095829588807e9*a*pow(A,0.02) -
6.632866669184202e7*pow(a,2)*pow(A,0.02) +
696168.259655105*pow(a,3)*pow(A,0.02) -
2.5569983466924567e9*pow(A,0.024) -
1.6290850543382132e9*a*pow(A,0.024) +
5.092958095093381e7*pow(a,2)*pow(A,0.024) +
2.1598214519034122e10*exp(0.0001*q)*cos(0.5*p) +
6.878799148940136e8*a*exp(0.0001*q)*cos(0.5*p) +
2.711489634805889e8*pow(a,2)*exp(0.0001*q)*cos(0.5*p) -
1.6271845370129107e7*pow(a,3)*exp(0.0001*q)*cos(0.5*p) -
2.872030838775911e6*pow(a,4)*exp(0.0001*q)*cos(0.5*p) +
2195.176090498871*pow(a,5)*exp(0.0001*q)*cos(0.5*p) +
0.44454042795482834*pow(a,6)*exp(0.0001*q)*cos(0.5*p) -
3.7440440350468895e10*pow(A,0.004)*exp(0.0001*q)*cos(0.5*p) -
3.2640043752665334e9*a*pow(A,0.004)*exp(0.0001*q)*cos(0.5*p) -
4.2381512989737064e8*pow(a,2)*pow(A,0.004)*exp(0.0001*q)*cos(0.5*p) +
3.2522752918584436e7*pow(a,3)*pow(A,0.004)*exp(0.0001*q)*cos(0.5*p) +
8.55670945143955e6*pow(a,4)*pow(A,0.004)*exp(0.0001*q)*cos(0.5*p) -
4505.568843712124*pow(a,5)*pow(A,0.004)*exp(0.0001*q)*cos(0.5*p) -
0.3720210247903439*pow(a,6)*pow(A,0.004)*exp(0.0001*q)*cos(0.5*p) -
3.0425389486546617e9*pow(A,0.008)*exp(0.0001*q)*cos(0.5*p) +
1.3827653653413448e9*a*pow(A,0.008)*exp(0.0001*q)*cos(0.5*p) -
1.1088009755144408e8*pow(a,2)*pow(A,0.008)*exp(0.0001*q)*cos(0.5*p) -
5.189438668636523e6*pow(a,3)*pow(A,0.008)*exp(0.0001*q)*cos(0.5*p) -
8.491167722970339e6*pow(a,4)*pow(A,0.008)*exp(0.0001*q)*cos(0.5*p) +
2305.3086535306234*pow(a,5)*pow(A,0.008)*exp(0.0001*q)*cos(0.5*p) +
2.4677949671362606e10*pow(A,0.012)*exp(0.0001*q)*cos(0.5*p) +
3.5102262262492533e9*a*pow(A,0.012)*exp(0.0001*q)*cos(0.5*p) +

2.3401121599709707e8*pow(a,2)*pow(A,0.012)*exp(0.0001*q)*cos(0.5*p) -
2.2316781098835904e7*pow(a,3)*pow(A,0.012)*exp(0.0001*q)*cos(0.5*p) +
2.8066292924911575e6*pow(a,4)*pow(A,0.012)*exp(0.0001*q)*cos(0.5*p) +
9.921083402133244e9*pow(A,0.016)*exp(0.0001*q)*cos(0.5*p) +
1.2470791471972056e8*a*pow(A,0.016)*exp(0.0001*q)*cos(0.5*p) +
2.1181426814501944e8*pow(a,2)*pow(A,0.016)*exp(0.0001*q)*cos(0.5*p) +
1.124912099861149e7*pow(a,3)*pow(A,0.016)*exp(0.0001*q)*cos(0.5*p) -
2.3271813423453487e10*pow(A,0.02)*exp(0.0001*q)*cos(0.5*p) -
4.309930003443114e9*a*pow(A,0.02)*exp(0.0001*q)*cos(0.5*p) -
1.8212632012157977e8*pow(a,2)*pow(A,0.02)*exp(0.0001*q)*cos(0.5*p) +
7.567153906801661 5e9*pow(A,0.024)*exp(0.0001*q)*cos(0.5*p) +
1.8660182790839946e9*a*pow(A,0.024)*exp(0.0001*q)*cos(0.5*p) -
1.7024304624291883e10*exp(0.0002*q)*pow(cos(0.5*p),2) +
1.0430772365755721e9*a*exp(0.0002*q)*pow(cos(0.5*p),2) -
3.7154874733179605e8*pow(a,2)*exp(0.0002*q)*pow(cos(0.5*p),2) +
7.225987102508308e7*pow(a,3)*exp(0.0002*q)*pow(cos(0.5*p),2) -
101180.63756725214*pow(a,4)*exp(0.0002*q)*pow(cos(0.5*p),2) +
116.45224641944915*pow(a,5)*exp(0.0002*q)*pow(cos(0.5*p),2) -
0.04359249422563403*pow(a,6)*exp(0.0002*q)*pow(cos(0.5*p),2) +
1.5622548889642405e10*pow(A,0.004)*exp(0.0002*q)*pow(cos(0.5*p),2) +
1.3336542362631638e9*a*pow(A,0.004)*exp(0.0002*q)*pow(cos(0.5*p),2) +
8.292994264835732e8*pow(a,2)*pow(A,0.004)*exp(0.0002*q)*pow(cos(0.5*p),2) -
2.0964349271891162e8*pow(a,3)*pow(A,0.004)*exp(0.0002*q)*pow(cos(0.5*p),2) +
195081.60150683302*pow(a,4)*pow(A,0.004)*exp(0.0002*q)*pow(cos(0.5*p),2) -

112.70527364740546*pow(a,5)*pow(A,0.004)*exp(0.0002*q)*pow(cos(0.5*p),2) +
1.2817363257312225e10*pow(A,0.008)*exp(0.0002*q)*pow(cos(0.5*p),2) -
2.4839494015423284e9*a*pow(A,0.008)*exp(0.0002*q)*pow(cos(0.5*p),2) -
3.0131706751120836e8*pow(a,2)*pow(A,0.008)*exp(0.0002*q)*pow(cos(0.5*p),2) +
2.0273223696731335e8*pow(a,3)*pow(A,0.008)*exp(0.0002*q)*pow(cos(0.5*p),2) -
94011.0109726146*pow(a,4)*pow(A,0.008)*exp(0.0002*q)*pow(cos(0.5*p),2)
-
9.721033240997374e8*pow(A,0.012)*exp(0.0002*q)*pow(cos(0.5*p),2) -
2.0689982822416382e9*a*pow(A,0.012)*exp(0.0002*q)*pow(cos(0.5*p),2) -
4.1694310619014955e8*pow(a,2)*pow(A,0.012)*exp(0.0002*q)*pow(cos(0.5*p),2) -
6.5335625434873424e7*pow(a,3)*pow(A,0.012)*exp(0.0002*q)*pow(cos(0.5*p),2) -
9.574664548141592e9*pow(A,0.016)*exp(0.0002*q)*pow(cos(0.5*p),2) +
1.5868480412934184e9*a*pow(A,0.016)*exp(0.0002*q)*pow(cos(0.5*p),2) +

2.6012902484012216e8*pow(a,2)*pow(A,0.016)*exp(0.0002*q)*pow(cos(0.5*p),2) -
7.375457222998259e9*pow(A,0.02)*exp(0.0002*q)*pow(cos(0.5*p),2) +
5.958776343267121e8*a*pow(A,0.02)*exp(0.0002*q)*pow(cos(0.5*p),2) +
6.480021120653882e9*pow(A,0.024)*exp(0.0002*q)*pow(cos(0.5*p),2) -
4.0835694714857845e9*exp(0.00030000000000000003*q)*pow(cos(0.5*p),3) -
2.95261373272097e9*a*exp(0.00030000000000000003*q)*pow(cos(0.5*p),3) -

```
3.6216642758879614e8*pow(a,2)*exp(0.00030000000000000003*q)*pow(cos(0.5*p),3) -
   510074.1061730366*pow(a,3)*exp(0.00030000000000000003*q)*pow(cos(0.5*p),3) +

191.18245348978263*pow(a,4)*exp(0.00030000000000000003*q)*pow(cos(0.5*p),3) -

0.09383021304018396*pow(a,5)*exp(0.00030000000000000003*q)*pow(cos(0.5*p),3) +

4.2814046711916176e10*pow(A,0.004)*exp(0.00030000000000000003*q)*pow(cos(0.5*p),3) +

2.605041797812924e9*a*pow(A,0.004)*exp(0.00030000000000000003*q)*pow(cos(0.5*p),3) +

1.0507670853991108e9*pow(a,2)*pow(A,0.004)*exp(0.00030000000000000003*q)*pow(cos(0.5*p),3) +

999166.3540880666*pow(a,3)*pow(A,0.004)*exp(0.00030000000000000003*q)*pow(cos(0.5*p),3) -

198.23769071920478*pow(a,4)*pow(A,0.004)*exp(0.00030000000000000003*q)*pow(cos(0.5*p),3) -

5.231056358777766e10*pow(A,0.008)*exp(0.00030000000000000003*q)*pow(cos(0.5*p),3) -

2.4722684433417654e9*a*pow(A,0.008)*exp(0.00030000000000000003*q)*pow(cos(0.5*p),3) -

9.917444437180276e8*pow(a,2)*pow(A,0.008)*exp(0.00030000000000000003*q)*pow(cos(0.5*p),3) -

504878.913812619*pow(a,3)*pow(A,0.008)*exp(0.00030000000000000003*q)*pow(cos(0.5*p),3) -

3.167676212146993e10*pow(A,0.012)*exp(0.00030000000000000003*q)*pow(cos(0.5*p),3) +

8.140155620890951e9*a*pow(A,0.012)*exp(0.00030000000000000003*q)*pow(cos(0.5*p),3) +

3.0361322475865906e8*pow(a,2)*pow(A,0.012)*exp(0.00030000000000000003*q)*pow(cos(0.5*p),3) +
```

7.221009703380188e10*pow(A,0.016)*exp(0.00030000000000000003*q)*pow(cos(0.5*p),3) -

5.32925497169427e9*a*pow(A,0.016)*exp(0.00030000000000000003*q)*pow(cos(0.5*p),3) -

2.6917714591202374e10*pow(A,0.02)*exp(0.00030000000000000003*q)*pow(cos(0.5*p),3) -
8.362710206715654e9*exp(0.0004*q)*pow(cos(0.5*p),4) +
4.795775645994299e9*a*exp(0.0004*q)*pow(cos(0.5*p),4) +
9.064637395114021e6*pow(a,2)*exp(0.0004*q)*pow(cos(0.5*p),4) +
162.243360605392*pow(a,3)*exp(0.0004*q)*pow(cos(0.5*p),4) +
5.002741554299767*pow(a,4)*exp(0.0004*q)*pow(cos(0.5*p),4) +
4.1959411741307014e10*pow(A,0.004)*exp(0.0004*q)*pow(cos(0.5*p),4) -
5.550750374061057e9*a*pow(A,0.004)*exp(0.0004*q)*pow(cos(0.5*p),4) -
3.5062819378000505e7*pow(a,2)*pow(A,0.004)*exp(0.0004*q)*pow(cos(0.5*p),4) +

11077.682150134939*pow(a,3)*pow(A,0.004)*exp(0.0004*q)*pow(cos(0.5*p),4) -
6.7067579002591354e10*pow(A,0.008)*exp(0.0004*q)*pow(cos(0.5*p),4) -
2.6858459647988596e9*a*pow(A,0.008)*exp(0.0004*q)*pow(cos(0.5*p),4) +

2.574567149547201e7*pow(a,2)*pow(A,0.008)*exp(0.0004*q)*pow(cos(0.5*p),4) +
4.0269164212116875e10*pow(A,0.012)*exp(0.0004*q)*pow(cos(0.5*p),4) +
3.4466456728000107e9*a*pow(A,0.012)*exp(0.0004*q)*pow(cos(0.5*p),4) -
6.819910500697979e9*pow(A,0.016)*exp(0.0004*q)*pow(cos(0.5*p),4) -
2.214061971328001e10*exp(0.0005*q)*pow(cos(0.5*p),5) -
2.460132760951408e9*a*exp(0.0005*q)*pow(cos(0.5*p),5) +
4.866888939333994e6*pow(a,2)*exp(0.0005*q)*pow(cos(0.5*p),5) -
3219.0158233049533*pow(a,3)*exp(0.0005*q)*pow(cos(0.5*p),5) +
4.150618963592535e10*pow(A,0.004)*exp(0.0005*q)*pow(cos(0.5*p),5) +
4.727277748721172e9*a*pow(A,0.004)*exp(0.0005*q)*pow(cos(0.5*p),5) -
4.867463528042816e6*pow(a,2)*pow(A,0.004)*exp(0.0005*q)*pow(cos(0.5*p),5) -
1.661556501433424e10*pow(A,0.008)*exp(0.0005*q)*pow(cos(0.5*p),5) -
2.268302866547495e9*a*pow(A,0.008)*exp(0.0005*q)*pow(cos(0.5*p),5) -
2.746972604510797e9*pow(A,0.012)*exp(0.0005*q)*pow(cos(0.5*p),5) +
1.0085503909026638e10*exp(0.0006000000000000001*q)*pow(cos(0.5*p),6) +
1.1028005204614408e7*a*exp(0.0006000000000000001*q)*pow(cos(0.5*p),6) +
34137.31412665328*pow(a,2)*exp(0.0006000000000000001*q)*pow(cos(0.5*p),6) -

1.9975666744725933e10*pow(A,0.004)*exp(0.0006000000000000001*q)*pow(cos(0.5*p),6) -

1.1257749511145892e7*a*pow(A,0.004)*exp(0.0006000000000000001*q)*pow(cos(0.5*p),6) +

```
9.891570568350595e9*pow(A,0.008)*exp(0.0006000000000000001*q)*pow(cos(0.5*p)
,6))  ;


        printf("Collision Efficiency %f", CE);

}
```

---

```
// C  program to calculate collision efficiency for particle range 0.4 to 2 µm
//// check in Turbo C++ (MS Windows (3.1))

#include <stdio.h>
#include <conio.h>
#include <math.h>
#include <stdlib.h>
#include <iostream.h>

void main()
{
//  p: Aerosol Density(in gm cm-3)
// a: Aerosol Radius(in micrometer)
// A: Drop Radius(in micrometer)
// q: Aerosol Charge(in e)
// CE: Collision Efficiency
double p;
double a;
double A;
double q;
double CE;
//printf("Please Enter the value of Aerosol density,Aerosol radius,Drop radius,Aerosol
charge respectively");
//scanf( "%4f %4f %4f %4f",&p,&a,&A,&q);
cout<<"Density of aerosol particle = ";
cin>>p;
cout<<"\n";
cout<<"Aerosol radius(0.4 to 2) = ";
cin>>a;
cout<<"\n";
cout<<"Drop radius = ";
cin>>A;
```

```
cout<<"\n";
cout<<"Aerosol charge = ";
cin>>q;
cout<<"\n";

    CE =exp(-2.0332482393118773e9 + 9.386397323856429e8*a -
5.368593346556073e8*pow(a,2) -
           3.8107987514284265e8*pow(a,3) + 1.0809512269713391e8*pow(a,4) -
3.681996905743367e7*pow(a,5) +
   142429.45223423408*pow(a,6) + 4.506330558121939e9*pow(A,0.004) -
2.0068584818287106e9*a*pow(A,0.004) +
   1.1520393452759135e9*pow(a,2)*pow(A,0.004) +
9.722442918019245e8*pow(a,3)*pow(A,0.004) -
   1.2209721352381226e8*pow(a,4)*pow(A,0.004) +
1.073212604257731e8*pow(a,5)*pow(A,0.004) -
   288511.54897538834*pow(a,6)*pow(A,0.004) -
5.923420248536376e8*pow(A,0.008) -
   5.225674957114702e8*a*pow(A,0.008) -
1.6843393841841048e8*pow(a,2)*pow(A,0.008) -
   4.8205204114087194e8*pow(a,3)*pow(A,0.008) -
2.6049796363097695e8*pow(a,4)*pow(A,0.008) -
   1.0420356071706103e8*pow(a,5)*pow(A,0.008) +
145898.6337429652*pow(a,6)*pow(A,0.008) -
   3.773509271921372e9*pow(A,0.012) + 1.3409868838963873e9*a*pow(A,0.012) -
   8.996967533556771e8*pow(a,2)*pow(A,0.012) -
6.091113836257848e8*pow(a,3)*pow(A,0.012) +
   4.5571668366031086e8*pow(a,4)*pow(A,0.012) +
3.37037623005092e7*pow(a,5)*pow(A,0.012) -
   8.400784197016721e8*pow(A,0.016) + 1.4419936264320617e9*a*pow(A,0.016) -
   1.2294177636742827e8*pow(a,2)*pow(A,0.016) +
6.803908798185091e8*pow(a,3)*pow(A,0.016) -
   1.8120558765895382e8*pow(a,4)*pow(A,0.016) +
3.9357786730623217e9*pow(A,0.02) -
   9.780912576445591e7*a*pow(A,0.02) +
1.0585332577032858e9*pow(a,2)*pow(A,0.02) -
   1.8039214637095708e8*pow(a,3)*pow(A,0.02) -
1.2235839919030554e9*pow(A,0.024) -
   1.0861266943054903e9*a*pow(A,0.024) -
4.8320031440245646e8*pow(a,2)*pow(A,0.024) +
   3.181117351269885e9*exp(0.0001*q)*cos(0.5*p) +
3.398035400453046e9*a*exp(0.0001*q)*cos(0.5*p) +
   3.4302480247160554e8*pow(a,2)*exp(0.0001*q)*cos(0.5*p) -
   5.3272685075580287e8*pow(a,3)*exp(0.0001*q)*cos(0.5*p) -
   8.267013575355259e7*pow(a,4)*exp(0.0001*q)*cos(0.5*p) +
   418149.2186832295*pow(a,5)*exp(0.0001*q)*cos(0.5*p) +
   10362.766361178523*pow(a,6)*exp(0.0001*q)*cos(0.5*p) -
```

5.561261602943977e9*pow(A,0.004)*exp(0.0001*q)*cos(0.5*p) -
2.491908636386257e9*a*pow(A,0.004)*exp(0.0001*q)*cos(0.5*p) +
4.295191682612865e8*pow(a,2)*pow(A,0.004)*exp(0.0001*q)*cos(0.5*p) +
1.2721372023825998e9*pow(a,3)*pow(A,0.004)*exp(0.0001*q)*cos(0.5*p) +
2.340909139758147e8*pow(a,4)*pow(A,0.004)*exp(0.0001*q)*cos(0.5*p) -
877730.4854761576*pow(a,5)*pow(A,0.004)*exp(0.0001*q)*cos(0.5*p) -
9961.684110121856*pow(a,6)*pow(A,0.004)*exp(0.0001*q)*cos(0.5*p) +
8.720730230036201e8*pow(A,0.008)*exp(0.0001*q)*cos(0.5*p) +
2.901655527836573e8*a*pow(A,0.008)*exp(0.0001*q)*cos(0.5*p) -
1.1033987340283413e9*pow(a,2)*pow(A,0.008)*exp(0.0001*q)*cos(0.5*p) -
6.376501735298591e8*pow(a,3)*pow(A,0.008)*exp(0.0001*q)*cos(0.5*p) -
2.2253258159952283e8*pow(a,4)*pow(A,0.008)*exp(0.0001*q)*cos(0.5*p) +
454688.7697232328*pow(a,5)*pow(A,0.008)*exp(0.0001*q)*cos(0.5*p) +
5.38828049569139e9*pow(A,0.012)*exp(0.0001*q)*cos(0.5*p) +
4.991801723732657e8*a*pow(A,0.012)*exp(0.0001*q)*cos(0.5*p) -
5.713575671748638e8*pow(a,2)*pow(A,0.012)*exp(0.0001*q)*cos(0.5*p) -
4.139390675096915e8*pow(a,3)*pow(A,0.012)*exp(0.0001*q)*cos(0.5*p) +
7.105492696532506e7*pow(a,4)*pow(A,0.012)*exp(0.0001*q)*cos(0.5*p) +
1.8799917186657686e9*pow(A,0.016)*exp(0.0001*q)*cos(0.5*p) -
3.732112428477611e9*a*pow(A,0.016)*exp(0.0001*q)*cos(0.5*p) +
1.1540883797230353e9*pow(a,2)*pow(A,0.016)*exp(0.0001*q)*cos(0.5*p) +
3.1210373217925364e8*pow(a,3)*pow(A,0.016)*exp(0.0001*q)*cos(0.5*p) -
4.987642825456881e9*pow(A,0.02)*exp(0.0001*q)*cos(0.5*p) -
6.084774572525726e9*a*pow(A,0.02)*exp(0.0001*q)*cos(0.5*p) -
2.475405327890841e8*pow(a,2)*pow(A,0.02)*exp(0.0001*q)*cos(0.5*p) -
5.926719879241283e8*pow(A,0.024)*exp(0.0001*q)*cos(0.5*p) +
8.048143637640426e9*a*pow(A,0.024)*exp(0.0001*q)*cos(0.5*p) -
3.1309764273331175e9*exp(0.0002*q)*pow(cos(0.5*p),2) -
1.3446763575876694e10*a*exp(0.0002*q)*pow(cos(0.5*p),2) -
1.1941430593017228e9*pow(a,2)*exp(0.0002*q)*pow(cos(0.5*p),2) +
6.615467429883574e8*pow(a,3)*exp(0.0002*q)*pow(cos(0.5*p),2) +
5.948452766065898e6*pow(a,4)*exp(0.0002*q)*pow(cos(0.5*p),2) -
11429.764575577181*pow(a,5)*exp(0.0002*q)*pow(cos(0.5*p),2) -
154.85285863914535*pow(a,6)*exp(0.0002*q)*pow(cos(0.5*p),2) -
1.008246193344071e9*pow(A,0.004)*exp(0.0002*q)*pow(cos(0.5*p),2) +
2.4376363542051187e9*a*pow(A,0.004)*exp(0.0002*q)*pow(cos(0.5*p),2) +
1.29762869852675e9*pow(a,2)*pow(A,0.004)*exp(0.0002*q)*pow(cos(0.5*p),2) -
1.9559676292962198e9*pow(a,3)*pow(A,0.004)*exp(0.0002*q)*pow(cos(0.5*p),2) -
8.657907275349338e6*pow(a,4)*pow(A,0.004)*exp(0.0002*q)*pow(cos(0.5*p),2) +
15921.290894958414*pow(a,5)*pow(A,0.004)*exp(0.0002*q)*pow(cos(0.5*p),2) -
2.042000026440195e9*pow(A,0.008)*exp(0.0002*q)*pow(cos(0.5*p),2) +
9.64651975724638e9*a*pow(A,0.008)*exp(0.0002*q)*pow(cos(0.5*p),2) -
1.0030269872369097e9*pow(a,2)*pow(A,0.008)*exp(0.0002*q)*pow(cos(0.5*p),2) +
1.9415664372826757e9*pow(a,3)*pow(A,0.008)*exp(0.0002*q)*pow(cos(0.5*p),2) +
2.82161610318651e6*pow(a,4)*pow(A,0.008)*exp(0.0002*q)*pow(cos(0.5*p),2) -
2.218674668656054e9*pow(A,0.012)*exp(0.0002*q)*pow(cos(0.5*p),2) +

1.4733743630129572e10*a*pow(A,0.012)*exp(0.0002*q)*pow(cos(0.5*p),2) +
2.5687975983088217e9*pow(a,2)*pow(A,0.012)*exp(0.0002*q)*pow(cos(0.5*p),2) -
6.467952413095375e8*pow(a,3)*pow(A,0.012)*exp(0.0002*q)*pow(cos(0.5*p),2) +
6.946237167768948e7*pow(A,0.016)*exp(0.0002*q)*pow(cos(0.5*p),2) +
1.0305173149291878e10*a*pow(A,0.016)*exp(0.0002*q)*pow(cos(0.5*p),2) -
1.6830108304965796e9*pow(a,2)*pow(A,0.016)*exp(0.0002*q)*pow(cos(0.5*p),2) +
3.166612820100934e9*pow(A,0.02)*exp(0.0002*q)*pow(cos(0.5*p),2) -
2.340715465955545e10*a*pow(A,0.02)*exp(0.0002*q)*pow(cos(0.5*p),2) +
4.516186639584399e9*pow(A,0.024)*exp(0.0002*q)*pow(cos(0.5*p),2) +
4.199931942353224e9*exp(0.00030000000000000003*q)*pow(cos(0.5*p),3) +
3.545775805171076e10*a*exp(0.00030000000000000003*q)*pow(cos(0.5*p),3) +

7.067841422867596e8*pow(a,2)*exp(0.00030000000000000003*q)*pow(cos(0.5*p),3)
-

1.0900895799416931e7*pow(a,3)*exp(0.00030000000000000003*q)*pow(cos(0.5*p),3
) -

1.400998826714275e6*pow(a,4)*exp(0.00030000000000000003*q)*pow(cos(0.5*p),3)
-

1700.0869728796463*pow(a,5)*exp(0.00030000000000000003*q)*pow(cos(0.5*p),3) +

1.6898480436932537e10*pow(A,0.004)*exp(0.00030000000000000003*q)*pow(cos(0.
5*p),3) -

1.815707129847539e10*a*pow(A,0.004)*exp(0.00030000000000000003*q)*pow(cos(0
.5*p),3) +

1.7112023550503469e9*pow(a,2)*pow(A,0.004)*exp(0.00030000000000000003*q)*po
w(cos(0.5*p),3) +

2.5605491815697663e6*pow(a,3)*pow(A,0.004)*exp(0.00030000000000000003*q)*po
w(cos(0.5*p),3) +

1.3102944417231055e6*pow(a,4)*pow(A,0.004)*exp(0.00030000000000000003*q)*po
w(cos(0.5*p),3) -

6.005063086854181e9*pow(A,0.008)*exp(0.00030000000000000003*q)*pow(cos(0.5*
p),3) -

4.746273688040182e10*a*pow(A,0.008)*exp(0.00030000000000000003*q)*pow(cos(0
.5*p),3) -

5.180305408052779e9*pow(a,2)*pow(A,0.008)*exp(0.00030000000000000003*q)*pow
(cos(0.5*p),3) +

7.719963189726753e6*pow(a,3)*pow(A,0.008)*exp(0.00030000000000000003*q)*pow(cos(0.5*p),3) -

6.795913206448919e9*pow(A,0.012)*exp(0.00030000000000000003*q)*pow(cos(0.5*p),3) -

1.5031893139845137e10*a*pow(A,0.012)*exp(0.00030000000000000003*q)*pow(cos(0.5*p),3) +

2.7853446902867503e9*pow(a,2)*pow(A,0.012)*exp(0.00030000000000000003*q)*pow(cos(0.5*p),3) +

9.908608346072227e9*pow(A,0.016)*exp(0.00030000000000000003*q)*pow(cos(0.5*p),3) +

4.467006698916325e10*a*pow(A,0.016)*exp(0.00030000000000000003*q)*pow(cos(0.5*p),3) -

1.6972286540068563e10*pow(A,0.02)*exp(0.00030000000000000003*q)*pow(cos(0.5*p),3) -
1.6018987923538242e10*exp(0.0004*q)*pow(cos(0.5*p),4) -
4.766381735947885e10*a*exp(0.0004*q)*pow(cos(0.5*p),4) -
1.289848381953121e9*pow(a,2)*exp(0.0004*q)*pow(cos(0.5*p),4) +
8.594068151182456e6*pow(a,3)*exp(0.0004*q)*pow(cos(0.5*p),4) +
26439.178347126614*pow(a,4)*exp(0.0004*q)*pow(cos(0.5*p),4) +
4.0510907063028593e9*pow(A,0.004)*exp(0.0004*q)*pow(cos(0.5*p),4) +
6.4081905180166504e10*a*pow(A,0.004)*exp(0.0004*q)*pow(cos(0.5*p),4) +
2.3925151027414255e9*pow(a,2)*pow(A,0.004)*exp(0.0004*q)*pow(cos(0.5*p),4) -
8.128879681366437e6*pow(a,3)*pow(A,0.004)*exp(0.0004*q)*pow(cos(0.5*p),4) -
1.577967948103728e10*pow(A,0.008)*exp(0.0004*q)*pow(cos(0.5*p),4) +
4.024248873989836e10*a*pow(A,0.008)*exp(0.0004*q)*pow(cos(0.5*p),4) -
1.1242339599631207e9*pow(a,2)*pow(A,0.008)*exp(0.0004*q)*pow(cos(0.5*p),4) +
1.2792699252594048e10*pow(A,0.012)*exp(0.0004*q)*pow(cos(0.5*p),4) -
5.6090886208009e10*a*pow(A,0.012)*exp(0.0004*q)*pow(cos(0.5*p),4) +
1.3642991300952475e10*pow(A,0.016)*exp(0.0004*q)*pow(cos(0.5*p),4) +
1.1759727848394487e10*exp(0.0005*q)*pow(cos(0.5*p),5) +
2.2052057260472206e10*a*exp(0.0005*q)*pow(cos(0.5*p),5) +
3.271541075583537e7*pow(a,2)*exp(0.0005*q)*pow(cos(0.5*p),5) -
122674.32163054819*pow(a,3)*exp(0.0005*q)*pow(cos(0.5*p),5) +
6.634097275148968e9*pow(A,0.004)*exp(0.0005*q)*pow(cos(0.5*p),5) -
5.9229429072389084e10*a*pow(A,0.004)*exp(0.0005*q)*pow(cos(0.5*p),5) -
2.1901256584925734e7*pow(a,2)*pow(A,0.004)*exp(0.0005*q)*pow(cos(0.5*p),5) -
8.99672042871235e9*pow(A,0.008)*exp(0.0005*q)*pow(cos(0.5*p),5) +
3.6849238990372185e10*a*pow(A,0.008)*exp(0.0005*q)*pow(cos(0.5*p),5) -
8.658687975433746e9*pow(A,0.012)*exp(0.0005*q)*pow(cos(0.5*p),5) -

```
    6.577230415847864e9*exp(0.0006000000000000001*q)*pow(cos(0.5*p),6) +
    3.7180133127891245e9*a*exp(0.0006000000000000001*q)*pow(cos(0.5*p),6) -

2.2939335540390946e6*pow(a,2)*exp(0.0006000000000000001*q)*pow(cos(0.5*p),6)
+

3.9992296416649423e9*pow(A,0.004)*exp(0.0006000000000000001*q)*pow(cos(0.5*
p),6) -

3.6398102806070423e9*a*pow(A,0.004)*exp(0.0006000000000000001*q)*pow(cos(0.
5*p),6) +

2.406073561582538e9*pow(A,0.008)*exp(0.0006000000000000001*q)*pow(cos(0.5*p)
,6))   ;



        printf("Collision Efficiency %f", CE);

}
```